\documentclass[fleqn,10pt]{wlscirep}
\usepackage[utf8]{inputenc}
\usepackage[T1]{fontenc}
\usepackage{xcolor}
\usepackage[normalem]{ulem}

\title{Low Lattice Thermal Conductivity of a Two-Dimensional Phosphorene Oxide}

\author[1]{Seungjun Lee}
\author[1,+]{Seoung-Hun Kang}
\author[1,*]{Young-Kyun Kwon}

\affil[1]{Department of Physics and
       Research Institute for Basic Sciences,
       Kyung Hee University, Seoul, 02447, Korea}
\affil[+]{Current address: Korea Institute for Advanced Study (KIAS), Seoul 02455, Korea}
\affil[*]{ykkwon@khu.ac.kr}

\begin{abstract}
A fundamental understanding of the phonon transport mechanism is
important for optimizing the efficiency of thermoelectric devices. In
this study, we investigate the thermal transport properties of the
oxidized form of phosphorene called phosphorene oxide (PO) by solving
phonon Boltzmann transport equation based on first-principles density
functional theory. We reveal that PO exhibits a much lower thermal
conductivity (2.42$-$7.08~W/mK at 300~K) than its pristine
counterpart as well as other two-dimensional materials. To comprehend
the physical origin of such low thermal conductivity, we scrutinize
the contribution of each phonon branch to the thermal conductivity by
evaluating various mode-dependent quantities including Gr\"uneisen
parameters, anharmonic three-phonon scattering rate, phase space of
three-phonon scattering processes. Our results show that its flexible
puckered structure of PO leads to smaller sound velocities; its
broken-mirror symmetry allows more ZA phonon scattering; and the
relatively-free vibration of dangling oxygen atoms in PO gives rise to
additional scattering resulting in further reduction in the phonon
lifetime. These results can be verified by the fact that PO has larger
phase space for three-phonon processes than phosphorene. Furthermore 
we show that the thermal conductivity of PO can be optimized by
controlling its size or its phonon mean free path, indicating that PO
can be a promising candidate for low-dimensional thermoelectric
devices.
\end{abstract}

\begin{document}

\flushbottom
\maketitle

\thispagestyle{empty}


\section*{Introduction}
The thermoelectric effect is one of the promising next-generation
energy harvesting technologies and has been studied extensively in
recent decades.~\cite{{Snyder2008},{Curtarolo2013}} The efficiency of
a thermoelectric device can be expressed using a dimensionless
quantity, $ZT$ defined as
$$ZT=\frac{S^2\sigma T}{\lambda + \kappa},$$
where $S$, $\sigma$, $T$, $\lambda$, and $\kappa$ represent the
Seebeck coefficient, electrical conductivity, absolute temperature,
electron and lattice thermal conductivities, respectively. In order to
increase $ZT$ value, we need to increase the power factor $S^2\sigma$,
and at the same time to decrease thermal conductivities. Seebeck
coefficient $S$ is usually much higher in semiconductor than in metal,
since $S$ is strongly related to differential total electron density
of state (DOS), which is maximized near band gaps, where DOS changes
abruptly. On the other hand, it is, unfortunately, impossible to
adjust $\sigma$ high, and $\lambda$ low simultaneously to enhance the
thermoelectricity, because they are coupled to each other and governed
by the Wiedemann-Franz law. In semiconductors, the phononic
contribution of the thermal conductivity ($\kappa$) is much larger
than its electronic contribution ($\lambda$). Moreover, $\kappa$ is
partially decoupled from $\sigma$, and thus its reduction is a useful
strategy for high-performance thermoelectric materials. Recently, a
few studies proposed that introduction of particular defect structures
increases $\sigma/(\lambda+\kappa)$ since the thermal conductivity
decreases much more than its electrical counterpart, although the
defects usually reduce both.~\cite{{Kim2015},{Anno2017}} Another
interesting research showed that making low-dimensional structures
enhances thermoelectric power. In the low-dimensional materials, $S$
can be further increased due to its unique properties, such as sudden
changes in the DOS,~\cite{Mahan1996} and quantum confinement
effects.~\cite{{Hicks1993},{Hicks1993a}} Moreover, such structural
redesign may affect thermal conductivity. $\kappa$ can be lowered more
while keeping $\sigma$ by forming superlattice or nanocrystal
structures,~\cite{{Boukai2008},{Majumdar777}} because the mean free
path of phonon is longer than that of electron. To realize state-of-the-art
thermoelectric devices, therefore, one should design a quasi
low-dimensional or superlattice structure with a semiconductor, which
has a low lattice thermal conductivity. It is, for example, known that
the best thermoelectric materials are
Bi$_{2}$Sb$_{3}$/Sb$_{2}$Te$_{3}$
superlattice~\cite{Venkatasubramanian2001} and layered SnSe
structure,~\cite{Zhao2014} which exhibit the highest $ZT$ values. 

Phosphorene, which has recently attracted attentions in various fields
related to materials research,~\cite{{Liu2014},{Akhtar2017}} would fit
in a category satisfying conditions mentioned above. Due to its
puckered structure representing strong anisotropy, its mechanical, electrical and thermal properties shows remarkably different
directional behavior, especially along the armchair and zigzag
directions.~\cite{{YC2},{YC3},{Qin2015},{Zhu2014},{Jain2015},{Xu2015}}
There have been several reports showing that phosphorene has a high
$ZT$ value along the armchair direction.\cite{{gangchen},{Fei2014}}

However, the poor stability of phosphorene under ambient air
conditions has been a crucial problem for its various applications
including thermoelectric devices. Earlier studies reported that
phosphorene can be easily oxidized in air because  of the lone-pair
electrons in each phosphorus
atom.~\cite{{Koenig2014},{Ziletti2015},{Wang2015},{jejune}} Although
there were a few proposals to prevent phosphorene from being
oxidized,~\cite{Wood2014} they are not easily applicable, and thus it
is still difficult to use phosphorene as a thermoelectric device.
Another way to overcome the poor air stability of phosphorene is 
to use the oxidized structure itself as a device. 
It was reported that a particular structure of phosphorene oxide (PO) is a semiconductor with a direct band gap of 0.6$\sim$0.88~eV,~\cite{{Ziletti2015},{Wang2015},{jejune}} which is slightly smaller than that of its pristine counterpart, phosphorene. It was also reported that PO has a relatively small electron effective mass of 0.18~$m_e$,~\cite{Wang2015} which is similar to that of phosphorene along the armchair direction and much smaller than along the zigzag direction.
In addition, phosphorene oxide exhibits some interesting physical properties such as non-symmorphic phase protected band structure.~\cite{jejune} 

Moreover, PO satisfies the conditions
mentioned above for good thermoelectric materials. In this paper, we
report our investigation on the thermal transport properties of the
oxidized structure of phosphorene by solving the Boltzmann transport
equation based on first-principles density functional theory. PO shows
significantly lower lattice thermal conductivity than phosphorene.
Furthermore, we present our analysis of why such low lattice thermal
conductivity was observed in PO. Our analysis includes the evaluation
of the phonon relaxation time, Gr\"uneisen parameters, and phase space
of the three-phonon processes, all of which enable us to understand
the fundamental origin of the low thermal conductivity of PO.

\section*{Results and Discussion}

\begin{figure}[t]
\includegraphics[width=1.0\columnwidth]{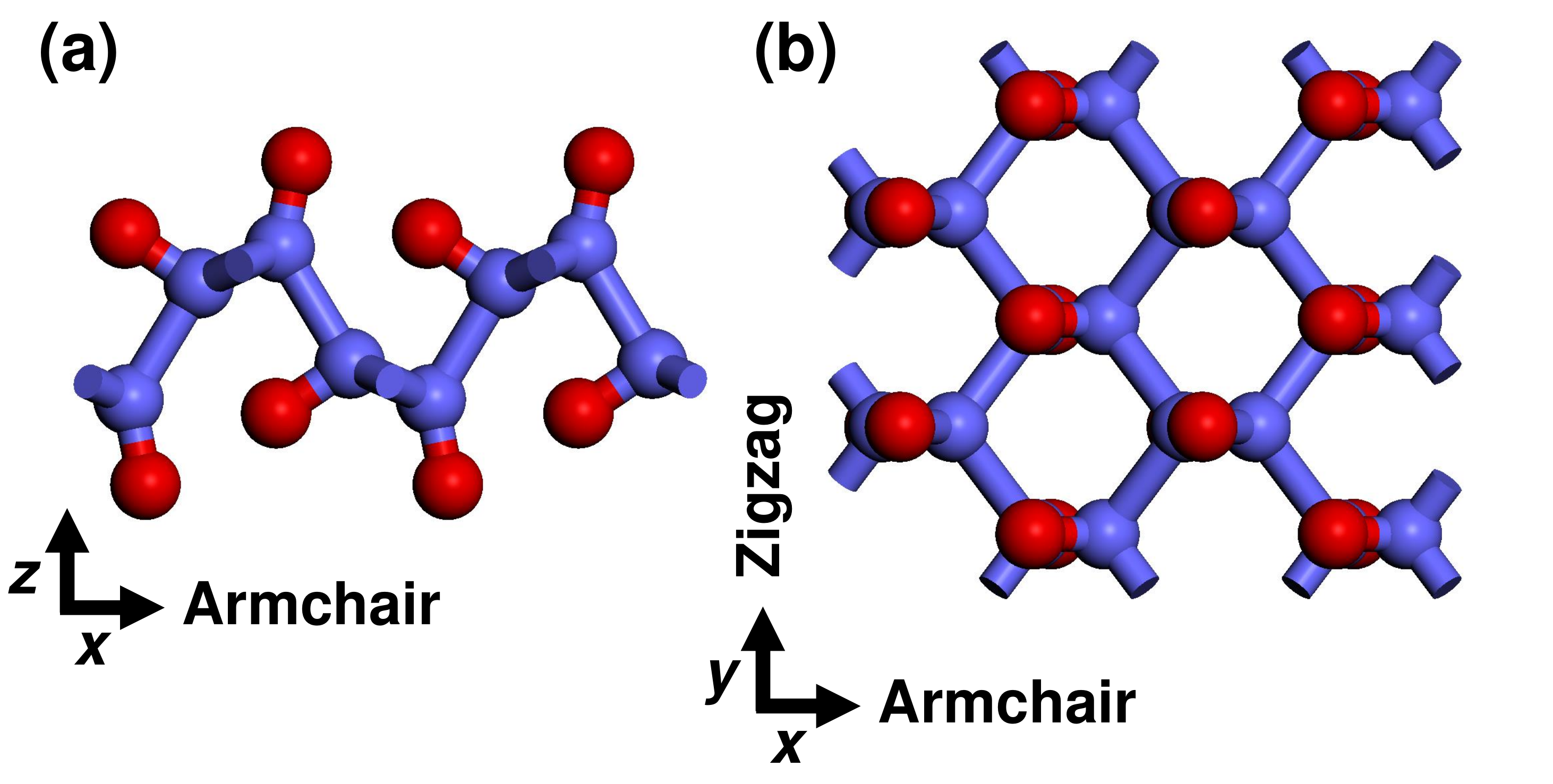}
\caption{Equilibrium structure of PO shown in side (a) and top (b)
views. Phosphorus and oxygen atoms are depicted by blue and red
spheres.
\label{structure}}
\end{figure}
In an earlier study~\cite{jejune}, 
We found the equilibrium
structure of PO, as shown in Fig.~\ref{structure}. Similar to
phosphorene, PO has a puckered configuration, composed of four
phosphorus and four oxygen atoms in a primitive unit cell. Every P
atom is connected to three other P atoms and one O atom, whereas every
O atom to one P atom and has three lone pairs of electrons. The
optimized lattice constants along the armchair and zigzag directions
were calculated to be 5.12 and 3.66~\AA, respectively, in good
agreement with previous studies of PO~\cite{{Wang2015},{Ziletti2015}}.
We expanded this equilibrium structure to $4\times4\times1$ super cell
to investigate the thermal properties.

\begin{figure}[t]
\includegraphics[width=1.0\columnwidth]{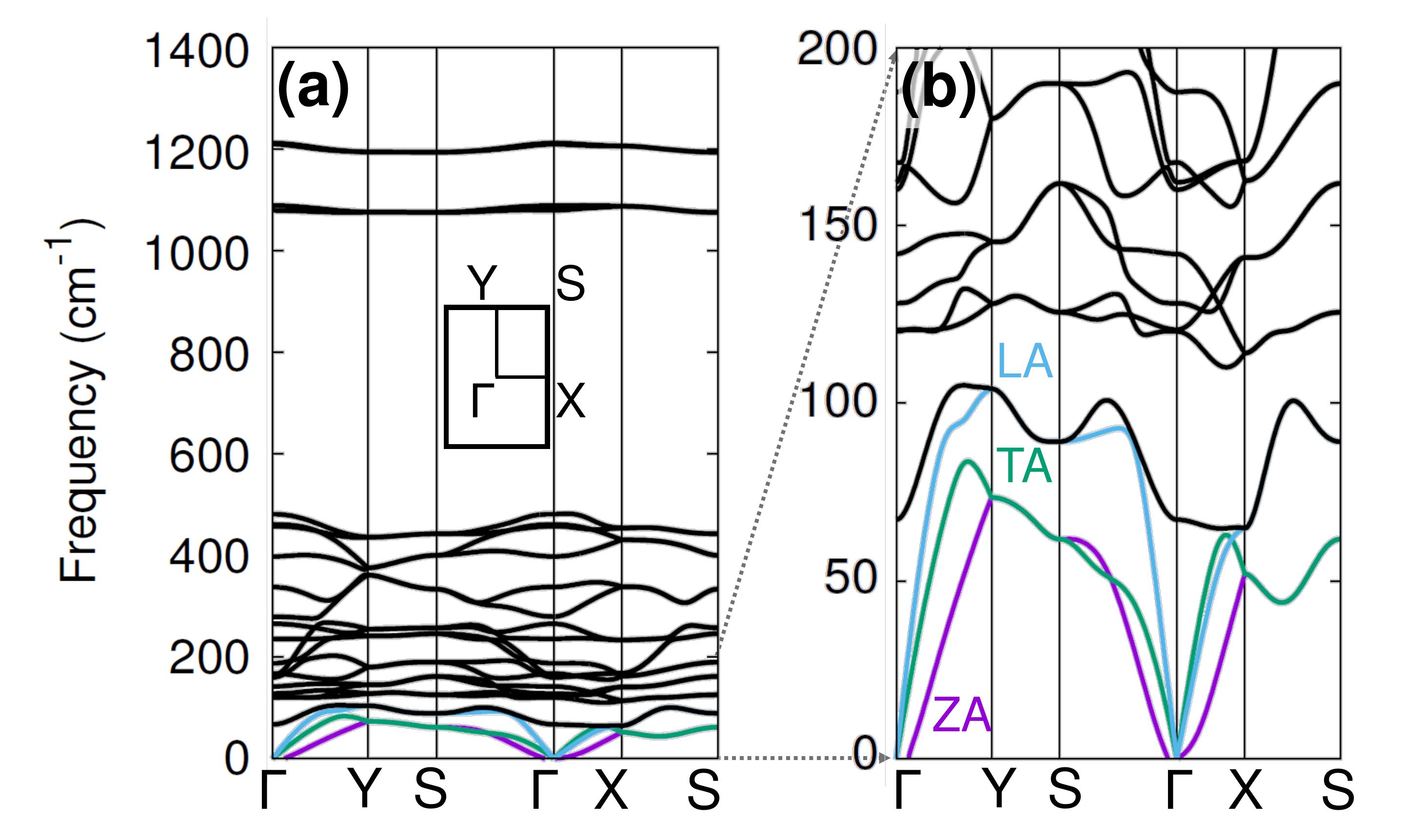}
\caption{Phonon dispersion of PO in (a) the whole frequency range and
in (b) the low frequency region ($\omega\le200$~cm$^{-1}$). The three
acoustic branches (LA: longitudinal, TA: transverse, ZA: flexural
modes) are represented respectively by sky blue, green, and purple
colors. The BZ of PO and the special path and points are shown in the
inset in (a).
\label{band}}
\end{figure}

\begin{figure}[t]
\includegraphics[width=1.0\columnwidth]{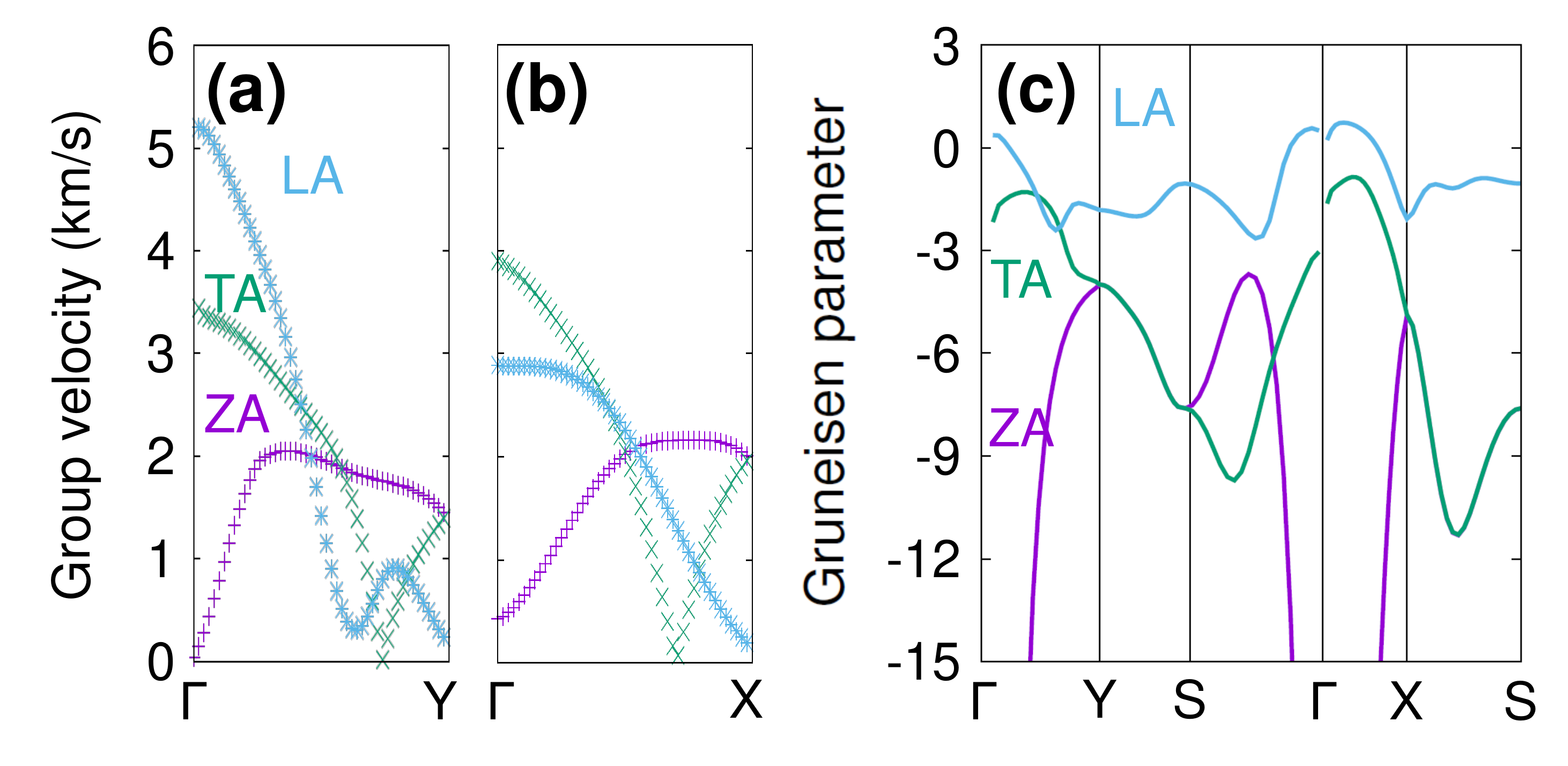}
\caption{Absolute values of phonon group velocities of PO for the three acoustic branches along (a) the $\Gamma$-Y and (b) the $\Gamma$-X directions. (c) Gr\"uneisen parameters calculated for the three acoustic modes in the first BZ. These calculated values for three modes LA, TA, and ZA are depicted by sky blue, green and purple colors, respectively. To compensate for the imaginary frequencies found in the ZA mode, which should exhibit the quadratic behavior near $\Gamma$, its group velocity along the $\Gamma$-Y direction was fitted using a polynomial function. 
\label{vgru}}
\end{figure}

Figure~\ref{band} shows our calculated phonon dispersion of PO
plotted along the special lines shown in its inset in (a). We focused
more on three acoustic branches (LA: longitudinal, TA: transverse, ZA:
flexural modes), which are main contribution to the
thermal conductivity. As shown in Fig.~\ref{band}(b), which is zoomed
in below $\omega\le200$ from (a), the acoustic modes exhibit
anisotropic dispersion along the $\Gamma$-Y (zigzag) and $\Gamma$-X
(armchair) directions. To clearly describe such an anisotropic
behaviors in the dispersion, we computed the group velocity $v_{g}$
of the three acoustic branches along these two directions. As shown in
Fig.~\ref{vgru} (a) and (b), the sound velocities of LA
and TA modes are in the range of $2.9-5.2$~km/s, which are slightly
smaller than those of other 2D materials such as silicene
($5.4-8.8$~km/s),~\cite{Li2013} MoS$_2$
($4.2-6.8$~km/s),~\cite{MoS2_Liu} graphene
($3.7-6.0$~km/s),~\cite{Ong2011} and phosphorene
($4.0-7.8$~km/s),~\cite{Qin2015} We found that $v_{g}$ of
the LA phonon mode along the zigzag direction is larger than that
along the armchair direction, as we easily understand since the LA
mode depends strongly on the elastic modulus of the material. Due to
the puckered structure shown in Fig.~\ref{structure} (a), PO can be
more easily stretched or compressed along the armchair direction
compared to along their zigzag counterpart, similar to phosphorene. On
the other hand, the group velocities of the TA mode do not show strong
directional dependence, and so do those of the ZA mode exhibiting the
quadratic behaviors as in planar graphene. Therefore, the LA mode
should be responsible for the anisotropicity in thermal transport
property of PO.

In a carrier transport phenomenon in a crystal solid, not only the
group velocity but the carrier lifetime or scattering rate also play
a crucial role in determining its transport properties. The evaluation
of the carrier lifetime requires anharmonic phonon-phonon scattering
or Umklapp process evolving three phonons, since the
momentum-conserved phonon scattering process described by only two
phonons is not relevant to the phonon lifetime.

We noticed that the
Gr\"uneisen parameter (GP), which is usually considered for the
thermal expansion behaviors, gives us a useful information on the
anharmonic phonon scattering rate. The mode-dependent GP is defined as
\begin{equation}
\label{equation2}
\gamma_j(\mathbf{q}) = -\frac{\partial\log\omega_j(\mathbf{q})}{\partial\log V},
\end{equation}
Figure~\ref{vgru} (c) shows
the GP values evaluated for the three acoustic phonon branches. The
ZA-mode GP $\gamma_{\textrm{ZA}}$ becomes very large in the long
wavelength region near the $\Gamma$ point as in other 2D materials~\cite{Sevik2014} as
described by membrane effect.~\cite{mem} It is also clear that
the GPs are largely anisotropic and discontinuous at the $\Gamma$
point, reflecting its structural features similar to the sound
velocity.
It was derived that the mode-dependent anharmonic phonon lifetime is
inversely proportional to the mode-dependent GP squared or
$\displaystyle{\tau_U(j,\mathbf{q})\propto\frac{1}{|\gamma_j(\mathbf{q})|^2}}$.~\cite{{klemens1},{klemens2},{Zou2001},{gru1-graphene}} Thus LA mode of PO,
which has smaller GP values than the TA or ZA, may exhibit longer
phonon lifetime than the other acoustic modes. Therefore, the LA mode
plays an important role in PO's thermal transport behavior as we will
later describe in detail.

\begin{figure}[t]
\includegraphics[width=1.0\columnwidth]{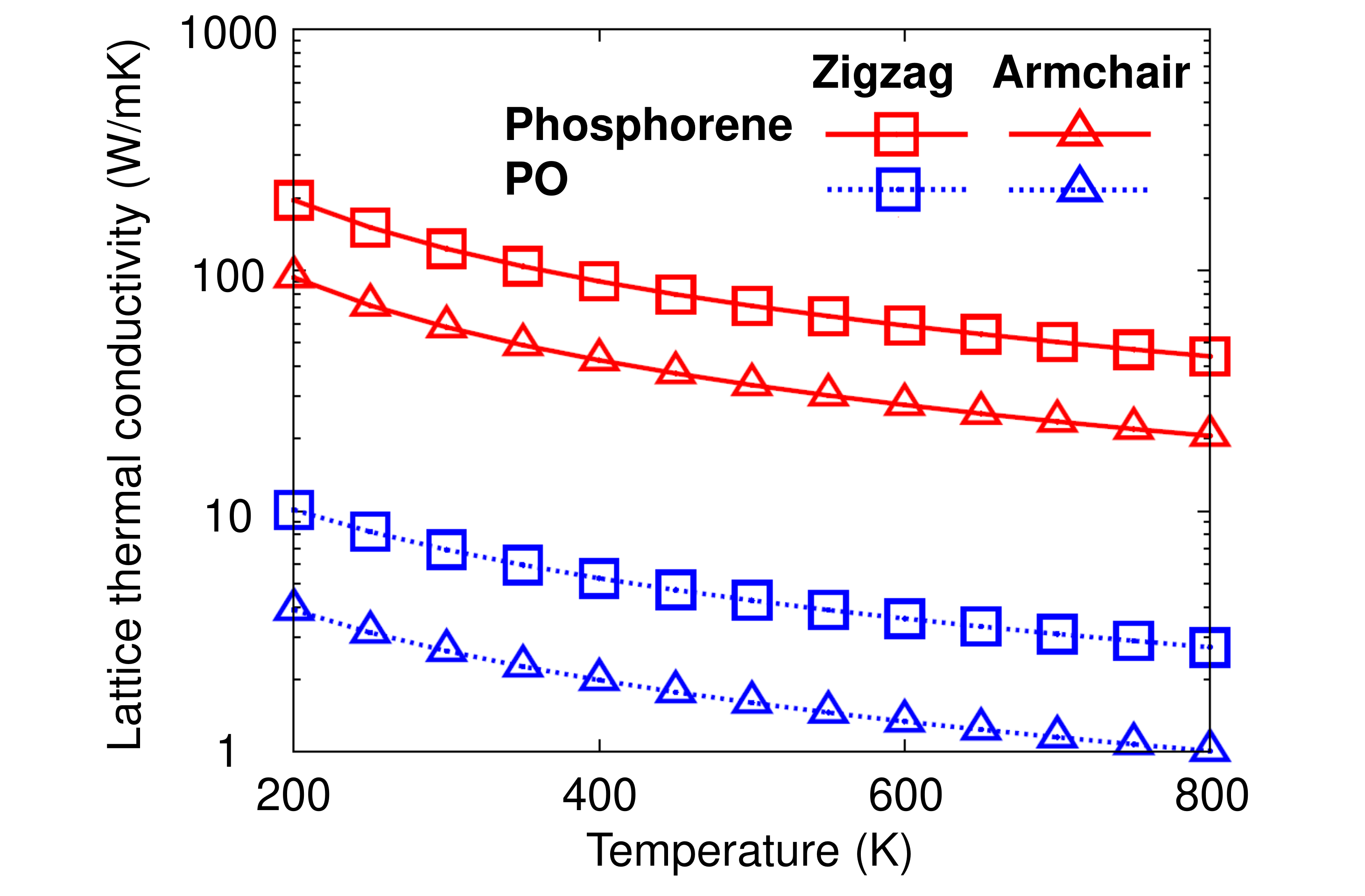}
\caption{Calculated lattice thermal conductivities of
phosphorene (red symbols) and PO (blue symbols) along zigzag (squares)
and armchair (triangles) directions as a function of temperature
ranging from 200~K to 800~K. The data points in each category were
well fitted with an inverse function $a/T$ with a single parameter
$a$, plotted with the solid or dashed line.
\label{kappa}}
\end{figure}

To investigate the temperature dependence of the lattice thermal 
conductivity $\kappa_{\mathrm{PO}}(T)$ of PO, we solved phonon
Boltzmann equation using the iterative approach mentioned in the
Method section. We also applied the same method to evaluate the
lattice thermal conductivity $\kappa_{\mathrm{P}}(T)$ of phosphorene
for comparison. Figure~\ref{kappa} shows
$\kappa^{\mathrm{A}}_{\mathrm{PO}}(T)$,
$\kappa^{\mathrm{Z}}_{\mathrm{PO}}(T)$, 
$\kappa^{\mathrm{A}}_{\mathrm{P}}(T)$, and
$\kappa^{\mathrm{Z}}_{\mathrm{P}}(T)$, where A and Z indicate the
armchair and zigzag directions along which the thermal conductivities
were calculated. These calculated values were almost perfectly fitted
with an inverse function $a/T$ with a single parameter $a$, which is
expected from the three-phonon anharmonic process.~\cite{Elseveir}
Our predicted room-temperature thermal conductivity values of phosphorene, which are 146~W/mK and 65~W/mK along zigzag and armchair directions, respectively, agree reasonably well with previously reported values ranging from 110 to 153~W/mK and from 33 to 64~W/mK along the respective directions~\cite{{Jain2015},{Xu2015},{Hong2015}}.
Both phosphorene and PO exhibit that their lattice thermal
conductivities along their armchair directions are approximately 2.5
times smaller than those along the zigzag directions, owing to the
structural anisotropy mentioned above. Although PO is structurally
similar to phosphorene, we observed that the thermal conductivity 
values of PO are much lower than those of phosphorene over the whole
temperature range regardless of the transport directions. For example,
at room temperature ($T=300$~K), we found
$\kappa^{\mathrm{A}}_{\mathrm{PO}}$ to be 2.42~W/mK, which is
lower than not only phosphorene
($\kappa^{\mathrm{A}}_{\mathrm{P}}=65$~W/mK), but also other 2D
materials such as silicene (26~W/mK),~\cite{doi:10.1063/1.4905540} 
MoS$_{2}$ (23.2$\sim$34.5~W/mK),~\cite{{YC1},{Peng2015},{Yan2014}} 
stanene (11.6~W/mK),~\cite{Peng2016} graphene
(3000$\sim$6000~W/mK),~\cite{Savas2000,Lindsay2011} and hexagonal
boron nitride (hBN) (350$\sim$600~W/mK).~\cite{bn2011,Lindsay2011} Low
thermal conductivity is one of sufficient conditions for
high-performance thermoelectric materials. In this sense, PO can be a new candidate for 2D thermoelectric materials.

\begin{table}[b]
\centering
\caption {Contribution of phonon branches in \% to the lattice thermal
conductivities of PO at 300~K along the zigzag and armchair
directions, $\kappa^{\mathrm{A}}_{\mathrm{PO}}$ and 
$\kappa^{\mathrm{Z}}_{\mathrm{PO}}$.
\label{table1}}
\begin{tabular}{c|cc} 
\hline
  &\multicolumn{2}{l}{Contribution of phonon branches in \% to}   \\
  & $\kappa^{\mathrm{A}}_{\mathrm{PO}}$ 
  & $\kappa^{\mathrm{Z}}_{\mathrm{PO}}$\\
\hline
  LA  & 42 & 27 \\
  TA  & 14 & 30 \\
  ZA  & 15 & 17 \\
  Others  & 29 & 26 \\
\hline
\end{tabular}
\end{table}

To explore what causes such a low lattice thermal conductivity in PO,
we first estimated the contributions of various phonon modes to the
$\kappa_\mathrm{PO}$ and summarized in Table~\ref{table1}. It was
revealed that the ZA phonon mode has a significantly long
lifetime in a completely flat 2D material because its mirror symmetry
limits the phase space for phonon-phonon scattering of the ZA phonon
mode.~\cite{Lindsay2010} For example, in graphene and hBN, whose
room-temperature thermal conductivities have been reported to be a few
thousands~\cite{Savas2000,Lindsay2010} and a few
hundreds~\cite{Lindsay2011,bn2011} in W/mK, their ZA modes
contribute to their thermal conductivities by about 
75~\%~\cite{Lindsay2010,Lindsay2011}. 
We found, on the other hand,
that the contribution of the ZA mode in PO becomes less significant to
be only $15\sim17$~\%, whereas the contribution from the other modes
including the LA and TA modes becomes dominant. This is because PO is
not a perfect 2D structure, but a puckered one, allowing more
phonon-phonon scattering of the ZA mode. Thus, puckered structures
would be advantageous for a low-dimensional thermoelectric
application.

\begin{figure}[t]
\includegraphics[width=1.0\columnwidth]{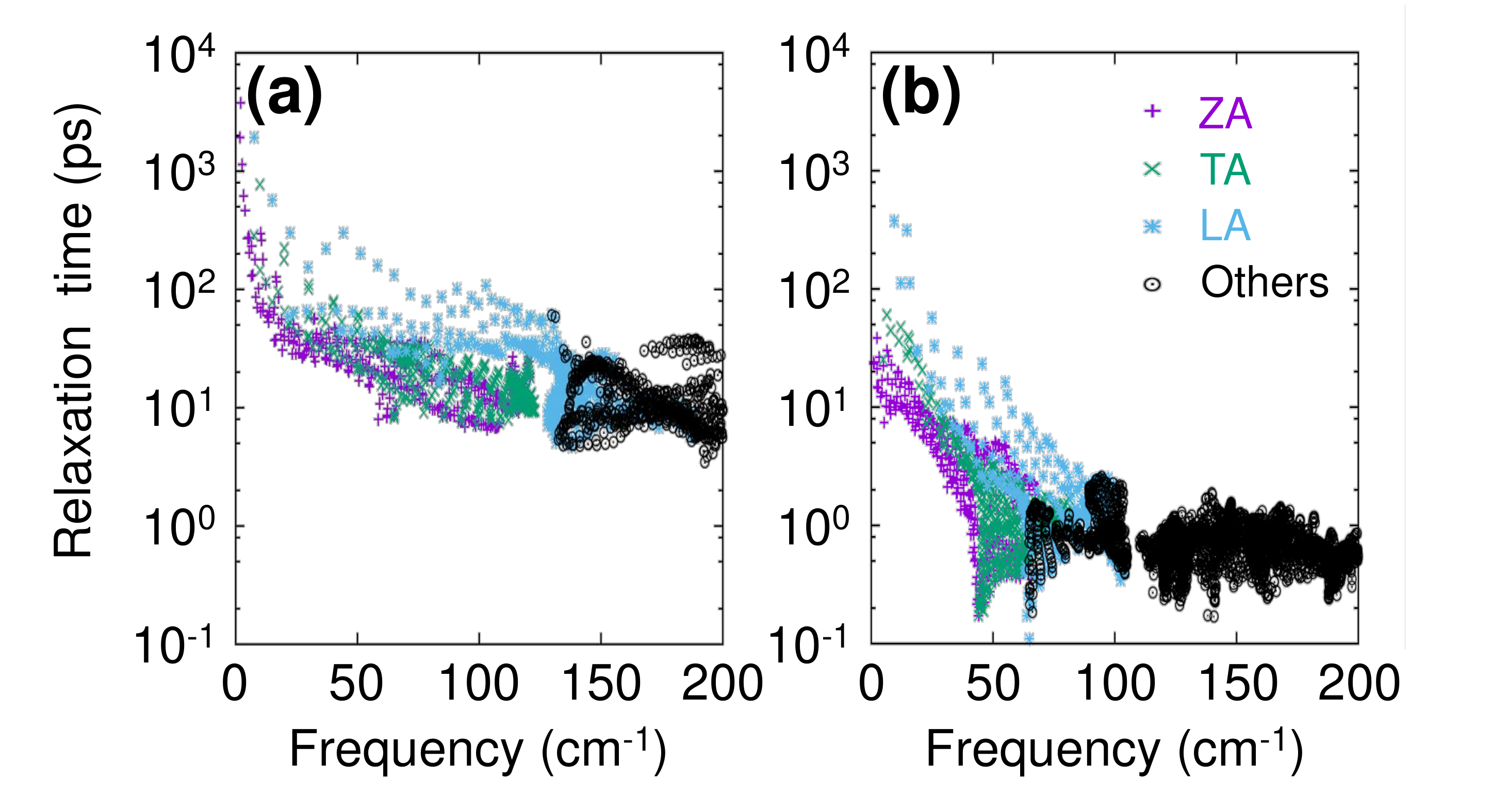}
\caption{The mode-dependent anharmonic phonon relaxation time of (a)
phosphorene and (b) PO for three acoustic modes, ZA (purple), TA
(green), LA (blue), and all the other modes (black). The relaxation
time is given in logarithmic scale.
\label{tau}}
\end{figure}

To further analyze the low lattice thermal conductivity of PO, we 
evaluated the relaxation times of phosphorene and PO as a function of
frequency. In general, there are various sources giving rise to phonon
scattering, for example, anharmonic phonon-phonon or Umklapp (U)
scattering, phonon-electron scattering, impurity effect, boundary
effect, isotope effect, and so on. Among these scattering sources, we
took only the phonon-phonon scattering into account, since the other
sources were much smaller in PO. Therefore, we simply replaced the
total scattering rate $1/\tau$ with the Umklapp scattering rate
$1/\tau_{\mathrm{U}}$. Figure~\ref{tau} shows our calculated
mode-dependent anharmonic phonon relaxation time of phosphorene and PO
for three acoustic modes (ZA, TA, and LA) and the other higher modes.
The phonon lifetime of PO is more than one-tenth smaller than that of
phosphorene as shown in Fig.~\ref{tau}. We assigned this reduction to
two effects of the oxygen atoms in PO. 
One is the trivial P-O composite effect, and the other is due to the
flexibility of oxygen atoms. Thus, oxidation occuring spontaneously on
the phosphorene surface, becomes an additional advantage leading to low
thermal conductivity. Each oxygen atom, which is connected only to one phosphrous atoms, is a kind of a dangling atom, whereas each phosphorous atom has four tetrahedral bonds with neighboring atoms. Thus, oxygen atoms may participate not only into their optical P$-$O vibration, but also vibrate along the in-plane directions together with phosphorous atoms contributing to the acoustic modes. This contribution may be responsible for the acoustic phonon softening leading to the reduction in the thermal conductivity of PO. To verify the effect of existence of such dangling atoms on the acoustic phonon modes, we devised a model structure mimicking the PO system. See Supplementary Information for our model calculation. 
\begin{figure}[t]
\includegraphics[width=1.0\columnwidth]{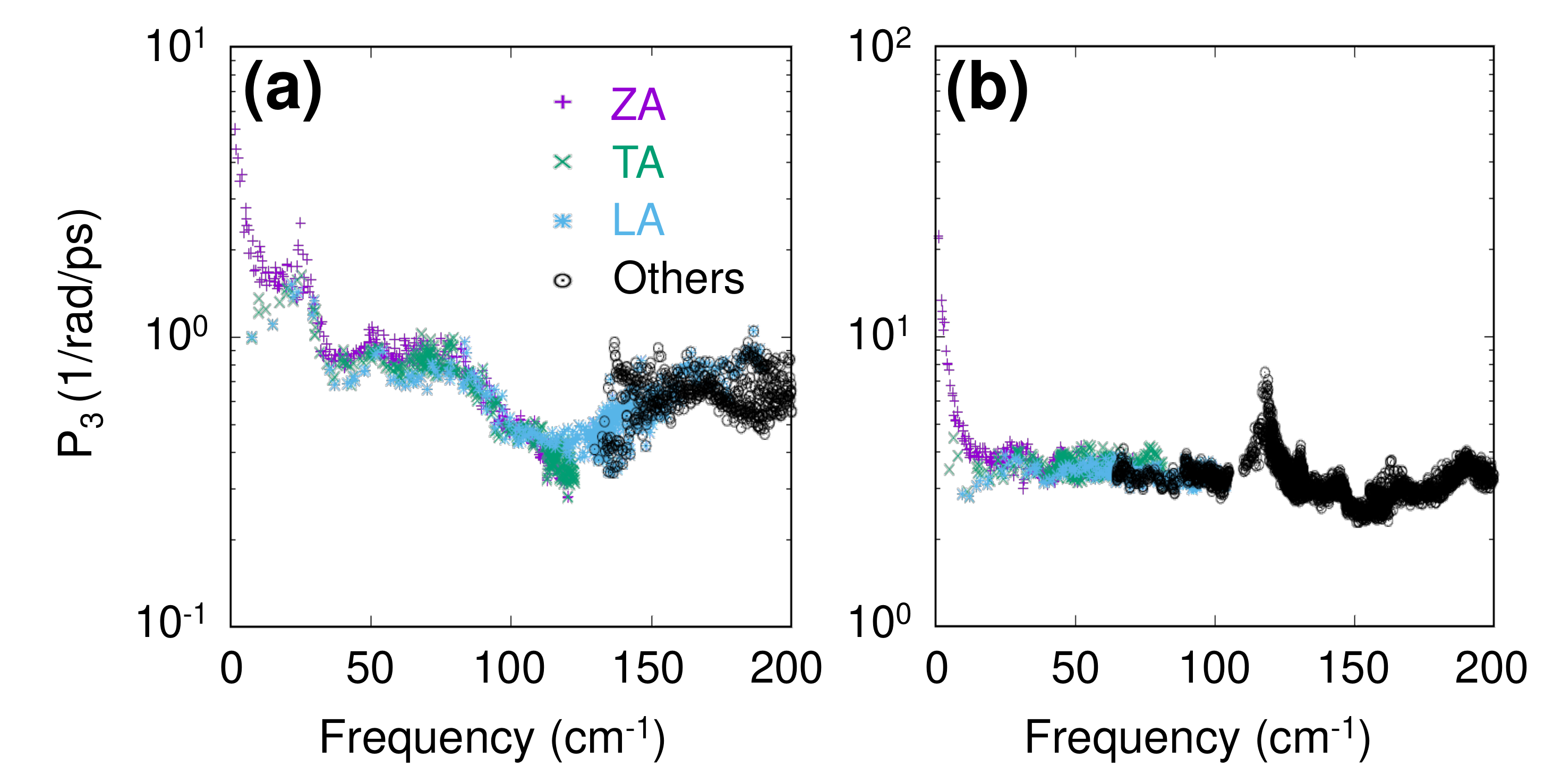}
\caption{Allowed mode-dependent phase spaces
$P_3(\omega_j(\mathbf{q}))$ of three-phonon scattering processes
evaluated in (a) phosphorene and (b) PO for three acoustic modes ZA
(purple), TA (green), LA (blue) and all the other modes (black). $P_3$
values are given in logarithmic scale.
\label{p3}}
\end{figure}

To discuss the physical phenomena of phonon scattering in more detail,
we analyzed the allowed phase space $P_3$ for the anharmonic
three-phonon scattering process introduced by Lindsay and
Broido.~\cite{P3original} 
The three-phonon scattering is allowed only when the energy and 
momentum conservation conditions are satisfied. 
Thus, the phonon scattering rate of each phonon state
is proportional to the number of available final states. In other
words, the phonon lifetime of each mode $\omega_j(\mathbf{q})$ is
inversely correlated with the mode-dependent phase space
$P_3(\omega_j(\mathbf{q}))$ defined by~\cite{Wu2016}
\begin{equation}
\label{equation3}
P_3(\omega_j(\mathbf{q}))= \frac{1}{\Omega_\mathrm{BZ}}\left[\frac{2}{3}D_j^{(+)}(\mathbf{q})+\frac{1}{3}D_j^{(-)}(\mathbf{q})\right],
\end{equation}
where  $\Omega_\mathrm{BZ}$ is the volume of BZ; 
$D_j^{(\pm)}(\mathbf{q})$ are two-phonon densities of
states~\cite{twoDOS} for absorption and emission processes,
respectively. Difference of a factor two between two coefficients was
introduced to avoid double counting. $D_j^{(\pm)}(\mathbf{q})$ can be
evaluated by
\begin{equation}
\label{equation5}
D_{j}^{(\pm)}(\mathbf{q}) = \sum_{j^\prime,j^{\prime\prime}}\int d\mathbf{q}^\prime\delta\left[\omega_{j}(\mathbf{q})\pm\omega_{j^\prime}(\mathbf{q}^\prime)-\omega_{j^{\prime\prime}}(\mathbf{q}\pm\mathbf{q}^\prime-\mathbf{G})\right].
\end{equation}
Here the momentum conservation was already imposed to replace
$\mathbf{q}^{\prime\prime}$ with
$\mathbf{q}\pm\mathbf{q}^\prime-\mathbf{G}$, where $\mathbf{G}$
vectors are the reciprocal lattice vectors to describe Umklapp
processes. Fig.~\ref{p3} shows $P_3$ values as a function of frequency
$\omega_j$ evaluated over the Brillouin zone. It is clear that PO has
a much larger P$_{3}$ value than phosphorene, indicating that more
scattering processes occur in PO than in phosphorene. This explains
why PO has a much shorter relaxation time than phosphorene.

\begin{figure}[t]
\includegraphics[width=1.0\columnwidth]{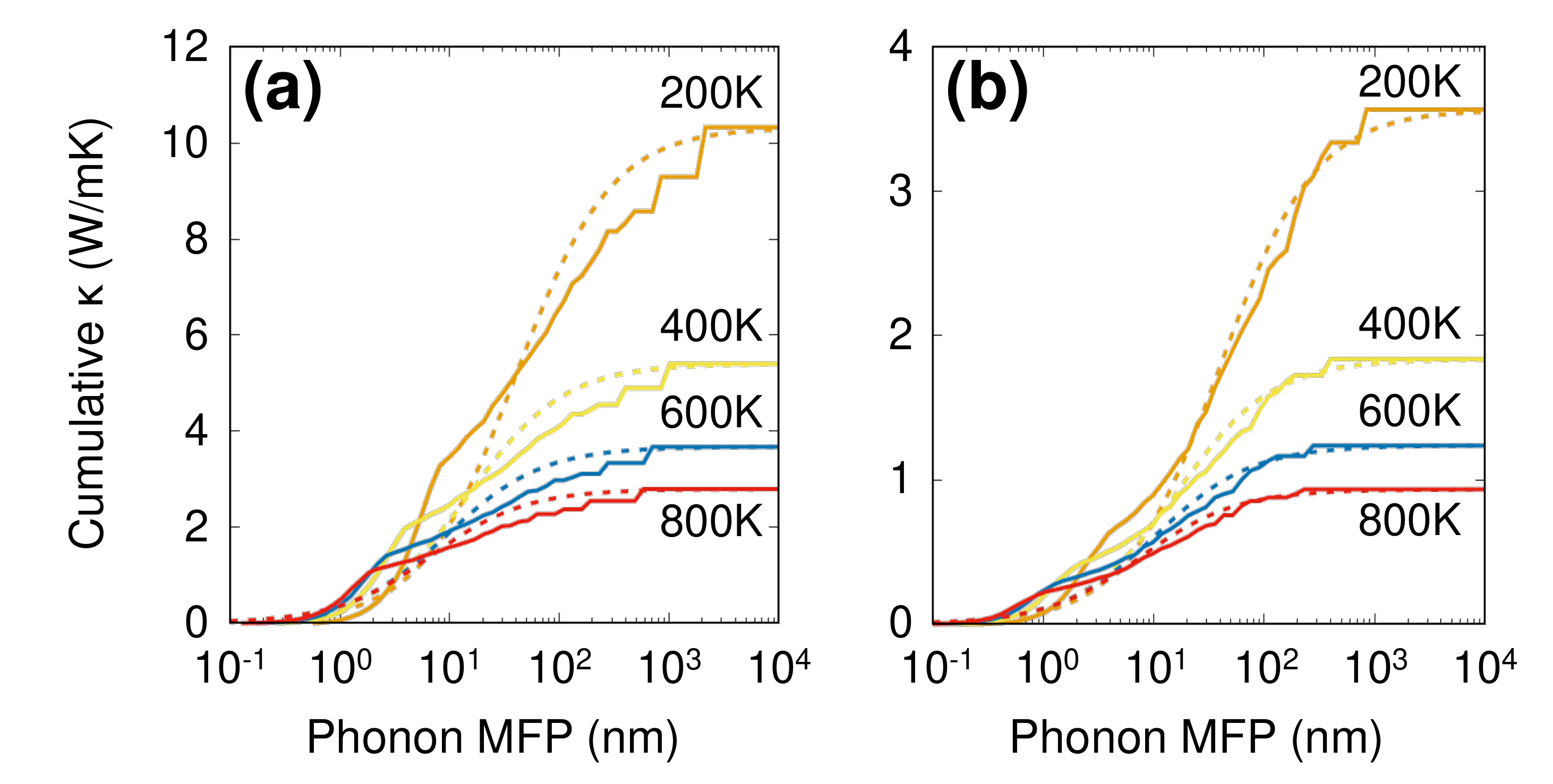}
\caption{Cumulative lattice thermal conductivities (solid lines) of PO
along (a) zigzag and (b) armchair directions with respect to the
phonon mean free path (MFP) at various temperatures. These
conductivities were fitted by the function defined in the text,
plotted with dotted lines.  
\label{cumulative_kappa}}
\end{figure}

To design high-performance thermoelectric devices, most important
strategy is to reduce the thermal conductivity while remaining its
electronic counterpart. Usually the lattice thermal conductivity of a
system can be decreased by reducing the size of the system, since the
size reduction results in shortening of its phonon MFP. On the other
hand, electron MFP is much smaller than that of phonon especially in
semiconductors, for example, the electron MFP of phosphorene is known
to be around only a few nm at 300K.~\cite{gangchen} Therefore,
optimization of grain size in nanocrystalline structures or nanowire
approach can be an effective way to enhance the thermoelectric
figure of merit. To explore the size dependence of thermal
conductivity of PO, we evaluated the cumulative thermal conductivity
$\kappa(l)$ by summing up all the contributions from phonon modes with
MFP smaller than $l$. Fig.~\ref{cumulative_kappa} shows the computed
$\kappa(l)$ as a function of phonon MFP along both zigzag and armchair
directions at various temperatures. As shown in the figure,
$\kappa(l)$ curves resemble a logistic function when MFP is given in
logarithmic scale, and thus we fit our calculated data to a single
parametric function given as~\cite{ShengBTE_2014} 
\begin{equation}
\label{equation6}
\kappa(l)=\frac{\kappa}{1+l_{0}/l},
\end{equation}
where $l_{0}$ is the fitting parameter determining the characteristic
value of the MFP, yielding $\kappa(l=l_0)=\kappa/2$. The $l_0$ values
are fitted to be 40.26 (39.98), 15.30 (16.46), 9.44 (10.48), and 6.71
(7.64)~nm at 200, 400, 600, and 800~K, respectively, for the zigzag
(armchair) direction.

\section*{Summary and Conclusions}

We presented our investigation on the thermal transport properties of
phosphorene oxide (PO) as well as phosphorene using the
first-principles calculations combining with the semiclassical
Boltzmann transport theory. We found that the thermal conductivity of
PO is much lower than those of other two-dimensional materials
including phosphorene, revealing that oxidation is responsible for the
reduction in the thermal conductivity. The thermal conductivity of PO
depends strongly on its transport directions, and were calculated to
be 2.42 and 7.08~W/mK along the armchair and the zigzag directions at
300~K. Similar to phosphorene, PO is structurally characterized by the
flexible puckered structure, which leads to lower-frequency acoustic
phonon modes and smaller sound velocities than genuine 2D flat
materials. In addition, PO possesses no mirror symmetry allowing more
ZA phonon scattering. Furthermore, we identified that nearly-free
vibration of dangling oxygen atoms gives rise to additional scattering
resulting in further reduction in the phonon lifetime. Spontaneous
oxidation of phosphorene greatly reduces its thermal conductivity,
which can be additionally optimized by controlling the size, and thus
PO can be a promising candidate for use in low-dimensional
thermoelectric devices.

\section*{Methods}

To investigate the thermal property of PO, we performed
first-principles calculations based on density functional
theory~\cite{Kohn1965} as implemented in the Vienna \textit{ab initio}
simulation package (VASP)~\cite{{Kresse1996},{Kresse1993}}. We
employed the projector augmented wave
potentials~\cite{{Blochl1994},{Kresse1999}} to describe the valence
electrons, and treated exchange-correlation functional within the
generalized gradient approximation of
Perdew-Burke-Ernzerhof~\cite{Perdew1996}. The plane-wave kinetic
energy cutoff was selected to be 500~eV, and $c=20$~{\AA} was chosen
for the lattice constant of the direction perpendicular to the plane
to minimize the interlayer interaction. The Brillouin zone (BZ) of
each structure was sampled using a $11{\times}11{\times}1$ $k$-point
grid for the primitive unit cells of phosphorene and PO. 

To precisely evaluate the temperature dependence of the lattice
thermal conductivity $\kappa_{\mathrm{PO}}(T)$ of PO, we solved the
phonon Boltzmann transport equation using an iterative approach
proposed by Omini~\textit{et al}~\cite{omini}, which implemented in
the ShengBTE code.~\cite{ShengBTE_2014} The thermal conductivity
tensor $\kappa^{\alpha\beta}$, where $\alpha$ and $\beta$ denote $x$,
$y$ or $z$, can be obtained from
\begin{equation}
\label{equation1}
\kappa^{\alpha\beta} = \frac{1}{k_{B}T^2\Omega N}\sum_{j,\mathbf{q}}f_{0}(f_{0}+1)(\hbar\omega_{j,\mathbf{q}})^2v_{j,\mathbf{q}}^{\alpha}L_{j,\mathbf{q}}^{\beta}
\end{equation}
where $k_{B}$ is the Boltzmann constant, $\Omega$ the unit cell
volume, and $N$ the $\Gamma$-centered $\mathbf{q}$-point grids. In the
summand, $f_{0}$ is the Bose-Einstein distribution function,
$\omega_{j,\mathbf{q}}$ and $\mathbf{v}_{j,\mathbf{q}}$ are phonon
frequency and group velocity of the phonon mode with the branch index
$j$ and wavevector $\mathbf{q}$. $\mathbf{L}_{j,\mathbf{q}}$ was
introduced to compensate the phonon distribution deviated from $f_0$
in the presence of the temperature gradient. This quantity with a
dimension of length can be expressed as
\begin{equation}
\label{eq-L}
\mathbf{L}_{j,\mathbf{q}}=\tau_{j,\mathbf{q}}^{0}(\mathbf{v}_{j,\mathbf{q}}+\mathbf{u}_{j,\mathbf{q}}),
\end{equation}
where $\tau_{j,\mathbf{q}}^{0}$ is the single-mode relaxation time
(SMRT) estimated from the imaginary part of self energy obtained by
many body perturbation theory including the anharmonic three phonon
scattering. Here, $\mathbf{u}_{j,\mathbf{q}}$ is the correction to the
SMRT approach due to the deviation of the phonon population at
the specific mode with $j$ and $\mathbf{q}$ from the Bose-Einstein
statistics. Since it is nonlinearly coupled with
$\mathbf{L}_{j,\mathbf{q}}$, we solved Eq.~(\ref{eq-L}) iteratively to
evaluate $\mathbf{u}_{j,\mathbf{q}}$ and thus
$\mathbf{L}_{j,\mathbf{q}}$. For more complete expressions of
$\tau_{j,\mathbf{q}}^{0}$, $\mathbf{L}_{j,\mathbf{q}}$, and
$\mathbf{u}_{j,\mathbf{q}}$, and their evaluation approach, see the
reference [\citenum{ShengBTE_2014}].
The phonon dispersion relations of phosphorene and PO were calculated by applying the finite displacement method~\cite{phonopy} to their respective $6{\times}6{\times}1$ supercells. We also computed their third-order interatomic force constants and phonon relaxation times to evaluate their correpsonding lattice thermal conductivities~\cite{{ShengBTE_2014},{PhysRevB.86.174307}} using $4{\times}4{\times}1$ supercell. We took up to the fourth nearest neighboring interaction and the corresponding BZ was sampled by a $2{\times}2{\times}1$ $k$-point grid.
Note that the cross-section area should be determined in order to
evaluate the thermal conductivity. As usually done in two dimensional
cases, the thicknesses of phosphorene and PO were set to 5.5 and
8.0~\AA,respectively, approximately corresponding to interlayer 
distances in their bulk configurations.

\section*{Data availability}
The datasets generated during and/or analysed during the current study
are available from the corresponding author on reasonable request.


\section*{Acknowledgments}
We gratefully acknowledge financial support from the Korean government
(MSIT) through the National Research Foundation (NRF) of Korea
(No. 2015R1A2A2A01006204). A portion of our computational work was
done using the resources of the KISTI Supercomputing Center
(KSC-2017-C2-0023, KSC-2018-C2-0033).

\section*{Author contributions statement}
S.L. performed all the calculations and analyzed the data with help from S.-H.K.and Y.-K.K.. S.L and Y.-K.K. wrote the manuscript, on which S.-H.K. commented. All authors reviewed the manuscript. Y.-K.K. supervised the whole project.

\section*{Competing interests}
The authors declare no competing interests.

\end{document}